\documentclass[pre,twocolumn,showpacs,superscriptaddress]{revtex4}
\usepackage{graphicx}
\usepackage{amsmath,amssymb}
\usepackage{listings,textcomp}

\newcommand{\av}[1]{\langle #1\rangle}
\newcommand{\mybigav}[1]{\Bigl\langle #1\Bigr\rangle}

\newcommand{\Peq}{P_{\mathrm{eq}}}

\newcommand{\dt}{\delta t}

\newcommand{\curly}[1]{{\mathbb #1}}
\newcommand{\curlyW}{\curly{W}}

\newcommand{\Eqref}[1]{Eq.~\eqref{#1}}
\newcommand{\Eqsref}[1]{Eqs.~\eqref{#1}}

\newcommand{\Line}[1]{{\itshape Line~#1}}
\newcommand{\Lines}[2]{{\itshape Lines~#1} and~{\itshape #2}}
\newcommand{\LineRange}[2]{{\itshape Lines~#1--#2}}

\newcommand{\MATLAB}{{\sc matlab}}
\newcommand{\Octave}{{\sc octave}}

\newcommand{\latin}[1]{{\itshape #1}}
\newcommand{\eg}{\latin{e.g.},}
\newcommand{\ie}{\latin{i.e.},}
\newcommand{\cf}{\latin{cf.}}

\newcommand{\etc}{\latin{etc.}}

\newcommand{\interalia}{\latin{inter alia}}

\begin{document}

\title{Malliavin weight sampling: a practical guide}

\author{Patrick B. Warren}

\affiliation{Unilever R\&D Port Sunlight, Quarry Road East, Bebington,
  Wirral, CH63 3JW, UK.}

\author{Rosalind J. Allen}

\affiliation{Scottish Universities Physics Alliance (SUPA), School of
  Physics and Astronomy, The University of Edinburgh, The Kings
  Buildings, Mayfield Road, Edinburgh, EH9 3JZ, UK}

\date{this version: December 28, 2013; published version:
{\it Entropy} {\bf16}, 221 (2014).}

\begin{abstract}
Malliavin weight sampling (MWS) is a stochastic calculus technique for
computing the derivatives of averaged system properties with respect
to parameters in stochastic simulations, without perturbing the
system's dynamics. It applies to systems in or out of equilibrium, in
steady state or time-dependent situations, and has applications in the
calculation of response coefficients, parameter sensitivities and
Jacobian matrices for gradient-based parameter optimisation
algorithms. The implementation of MWS has been described in the
specific contexts of kinetic Monte Carlo and Brownian dynamics
simulation algorithms. Here, we present a general theoretical
framework for deriving the appropriate MWS update rule for any
stochastic simulation algorithm. We also provide pedagogical
information on its practical implementation.
\end{abstract}

\pacs{%
05.10.-a, % Computational methods in statistical physics and nonlinear dynamics
05.40.-a, % Fluctuation phenomena, random processes, noise, and Brownian motion
05.70.Ln} %  Nonequilibrium and irreversible thermodynamics

\maketitle

\section{Introduction}
Malliavin weight sampling (MWS) is a method for computing derivatives
of averaged system properties with respect to parameters in stochastic
simulations \cite{Bel07, Nua06}. The method has been used in
quantitative financial modelling to obtain the ``Greeks'' (price
sensitivities) \cite{FLL+99}; and, as the Girsanov transform, in
kinetic Monte Carlo simulations for systems biology \cite{PA07}.
Similar ideas have been used to study fluctuation-dissipation
relations in supercooled liquids \cite{Ber07}. However, MWS appears to
be relatively unknown in the fields of soft matter, chemical and
biological physics, perhaps because the theory is relatively
impenetrable for non-specialists, being couched in the language of
abstract mathematics (\eg\ martingales, Girsanov transform, Malliavin
calculus, \etc); an exception in financial modelling is
Ref.~\cite{CG07}.

MWS works by introducing an auxiliary stochastic quantity, the
Malliavin weight, for each parameter of interest. The Malliavin
weights are updated alongside the system's usual (unperturbed)
dynamics, according to a set of rules. The derivative of any system
function, $A$, with respect to a parameter of interest is then given
by the average of the product of $A$ with the relevant Malliavin
weight, or in other words by a weighted average of $A$, in which the
weight function is given by the Malliavin weight. Importantly, MWS
works for non-equilibrium situations, such as time-dependent processes
or driven steady states. It thus complements existing methods based on
equilibrium statistical mechanics, which are widely used in soft
matter and chemical physics.

MWS has so far been discussed only in the context of specific
simulation algorithms. In this paper, we present a pedagogical and
generic approach to the construction of Malliavin weights, which can
be applied to any stochastic simulation scheme. We further describe
its practical implementation in some detail using as our example one
dimensional Brownian motion in a force field.

\section{The construction of Malliavin weights}
The rules for the propagation of Malliavin weights have been derived
for the kinetic Monte-Carlo algorithm \cite{PA07, WA12}, for the
Metropolis Monte-Carlo scheme \cite{Ber07} and for both underdamped
and overdamped Brownian dynamics \cite{WA12b}. Here, we present a
generic theoretical framework, which encompasses these algorithms and
also allows extension to other stochastic simulation schemes.

We suppose that our system evolves in some state space, and a point in
this state space is denoted as $S$. Here, we assume that the state space
is continuous, but our approach can easily be
translated to discrete or mixed discrete-continuous state spaces.
Since the system is stochastic, its state at time $t$ is described by
a probability distribution, $P(S)$. In each simulation step, the state
of the system changes according to a propagator, $W(S\to S')$, which
gives the probability that the system moves from point $S$ to point
$S'$ during an application of the update algorithm. The propagator
has the property that
\begin{equation}
P'(S')=\int_S\!dS\,\, W(S\to S')\,P(S)
\label{eq:prop1}
\end{equation}
where $P'(S)$ is the probability distribution after the update step
has been applied and the integral is over the whole state space. We
shall write this in a shorthand notation as
\begin{equation}
P'=\int WP\,.
\label{eq:prop2}
\end{equation}
Integrating \Eqref{eq:prop1} over $S'$, we see that the propagator
must obey $\int_{S'}W(S\to S')=1$. It is important to note, however,
that we do \emph{not} assume the detailed balance condition,
$\Peq(S)\,W(S\to S')=\Peq(S')\,W(S'\to S)$ for some equilibrium
$\Peq(S)$. Thus, our results apply to systems whose dynamical rules do
not obey detailed balance (such as chemical models of gene regulatory
networks \cite{WtW05}), as well as to systems out of steady state. 

We observe that the (finite) product
\begin{equation}
\begin{split}
\curlyW(S_1,\dots, S_n) &= 
W(S_1\to S_2)\times\cdots\\
&{}\hspace{2em}{}\times W(S_{n-1}\to S_n)
\end{split}
\label{eq:ptraj}
\end{equation} 
is proportional to the probability of occurrence of a trajectory of
states, $\{S_1,\dots, S_n\}$, and can be interpreted as a
\emph{trajectory weight}.

Let us now consider the average of some quantity, $A(S)$, over the state space,
in shorthand
\begin{equation}
\av{A}=\int\! A\,P\,.
\end{equation}
The quantity, $A$, might well be a complicated function of the state of
the system: for example the extent of crystalline order in a
particle-based simulation, or a combination of the concentrations of
various chemical species in a simulation of a biochemical network. We
suppose that we are interested in the sensitivity of $\av{A}$ to
variations in some parameter of the simulation, which we denote as
$\lambda$. This might be one of the force field parameters (or the
temperature) in a particle-based simulation or a rate constant in a
kinetic Monte Carlo simulation. We are interested in computing
$\partial\av{A}/\partial\lambda$. This quantity can be written as
\begin{equation}
\frac{\partial\av{A}}{\partial\lambda} =
\int\! A\,\frac{\partial P}{\partial\lambda} =
\int\! A\,P\,Q_\lambda \,
\label{eq6}
\end{equation}
where
\begin{equation}
Q_\lambda = \frac{\partial\ln P}{\partial\lambda}\,.
\end{equation}
Let us now suppose that we track in our simulation not only the
physical state of the system, but also an auxiliary stochastic variable,
which we term $q_\lambda$. At each simulation step, $q_\lambda$ is
updated according to a rule that depends on the system state; this
does not perturb the system's dynamics, but merely acts as a
``readout''. By tracking $q_\lambda$, we \emph{extend} the state space,
so that $S$ becomes $\{S,q_\lambda\}$. We can then define the average
$\av{q_\lambda}_S$, which is an average of the value of $q_\lambda$
in the extended state space, with the constraint that the original
(physical) state space point is fixed at $S$ (see further below).

Our aim is to define a set of rules for updating $q_\lambda$, such that
$\av{q_\lambda}_S=Q_\lambda$, \ie\  such that the average of the
auxiliary variable, for a particular state space point, measures the
{\em{derivative}} of the probability distribution with respect to the
parameter of interest, $\lambda$. If this is the case then, from
\Eqref{eq6}
\begin{equation}
\frac{\partial\av{A}}{\partial\lambda}=\av{A\,q_\lambda}\,.
\label{eq2}
\end{equation}
The auxiliary variable, $q_\lambda$, is the Malliavin weight
corresponding to the parameter, $\lambda$.

How do we go about finding the correct updating rule? If the
Malliavin weight exists, we should be able to derive its updating rule
from the system's underlying stochastic equations of motion. We
obtain an important clue from differentiating \Eqref{eq:prop1}
with respect to $\lambda$. Extending the shorthand notation, one finds
\begin{equation}
P'Q_\lambda'=\int WP\,
\Bigl(Q_\lambda+\frac{\partial\ln W}{\partial\lambda}\Bigr)\,.
\label{eq:prop3}
\end{equation}
This strongly suggests that the rule for updating the Malliavin weight
should be
\begin{equation}
q_\lambda'=q_\lambda+\frac{\partial\ln W}{\partial\lambda}\,.
\label{eq:prop4}
\end{equation}
In fact, this is correct. The proof is not difficult and, for the case
of Brownian dynamics, can be found in the supplementary material for
Ref.~\cite{WA12b}. It involves averaging \Eqref{eq:prop4} in the
extended state space,~$\{S,q_\lambda\}$.

From a practical point of view, for each time step, we implement the
following procedure:
\begin{itemize}
\item propagate the system from its current state, $S$, to a new state,
 $S'$, using the algorithm that implements the stochastic equations of
 motion (Brownian, kinetic Monte-Carlo, \etc);
\item with knowledge of $S$ and $S'$, and the
 propagator, $W(S\to S')$, calculate the change in the Malliavin
 weight $\Delta q_\lambda = \partial\ln W(S\to S')/\partial\lambda$;
\item update the Malliavin weight according to $q_\lambda\to
 q_\lambda'=q_\lambda + \Delta q_\lambda$.
\end{itemize}

At the start of the simulation, the Malliavin weight is usually
initialised to $q_\lambda=0$.

Let us first suppose that our system is not in steady state, but rather
the quantity $\av{A}$ in which we are interested is changing
in time, and likewise $\partial\av{A(t)}/\partial\lambda$ is a
time-dependent quantity. To compute
$\partial\av{A(t)}/\partial\lambda$, we run $N$ independent
simulations, in each one tracking as a function of time $A(t)$,
$q_{\lambda}(t)$ and the product, $A(t)\,q_{\lambda}(t)$. The
quantities $\av{A(t)}$ and $\partial\av{A(t)}/\partial\lambda$ are
then given by
\begin{subequations}
\label{eq:samp}
\begin{align}
&\av{A(t)}\approx \frac{1}{N}\sum_{i=1}^N A_i(t)\,,\\[6pt]
&\frac{\partial\av{A(t)}}{\partial\lambda}\approx 
\frac{1}{N}\sum_{i=1}^N A_i(t)\,q_{\lambda,i}(t)\,,
\end{align}
\end{subequations}
where $A_i(t)$ is the value of $A(t)$ recorded in the $i$th simulation
run (and likewise for $q_{\lambda,i}(t)$). Error estimates can be
obtained from replicate simulations.

If, instead, our system is in steady state, the procedure needs to be
modified slightly. This is because the variance in the values of
$q_{\lambda}(t)$ across replicate simulations increases linearly in
time (this point is discussed further below). For long times,
computation of $\partial\av{A}/\partial\lambda$ using
\Eqref{eq:samp} therefore incurs a large statistical
error. Fortunately, this problem can easily be solved, by computing
the correlation~function
\begin{equation}
C(t, t')=\av{A(t)\,[q_\lambda(t)-q_\lambda(t')]}\,.
\label{eq:c1}
\end{equation}
In steady state, $C(t,t') = C(t-t')$, with the property that $C(\Delta
t)\to\partial A/\partial\lambda$ as $\Delta t\to\infty$. In a single
simulation run, we simply measure $q_\lambda(t)$ and $A(t)$ at time
intervals separated by $\Delta t$ (which is typically multiple
simulation steps). At each measurement, we compute
$A(t)\,[q_\lambda(t)-q_\lambda(t-\Delta t)]$. We then average this
latter quantity over the whole simulation run to obtain an estimate of
$\partial\av{A}/\partial\lambda$. For this estimate to be accurate, we
require that $\Delta t$ is long enough that $C(\Delta t)$ has reached
its plateau value; this typically means that $\Delta t$ should be
longer than the typical relaxation time of the system's dynamics. The
correlation function approach is discussed in more detail in
Refs.~\cite{WA12, WA12b}.

Returning to a more theoretical perspective, it is interesting to note
that the rule for updating the Malliavin weight, \Eqref{eq:prop4},
depends deterministically on $S$ and $S'$. This implies that the value
of the Malliavin weight at time $t$ is completely determined by the
trajectory of system states during the time interval, $0 \to t$. In
fact, it is easy to show that
\begin{equation}
q_\lambda=\frac{\partial\ln\curlyW}{\partial\lambda}
\label{eq:qw}
\end{equation}
where $\curlyW$ is the trajectory weight defined in
\Eqref{eq:ptraj}. Similar expressions are given in
Refs.~\cite{Ber07, WA12}. Thus, the Malliavin weight, $q_\lambda$, is
not fixed by the state point, $S$, but by the entire trajectory of
states that have led to state point $S$. Since many different
trajectories can lead to $S$, many values of $q_\lambda$ are possible
for the same state point, $S$. The average $\av{q_\lambda(t)}_S$ is
actually the expectation value of the Malliavin weight, averaged over
all trajectories that reach state point $S$ at time $t$. This can be
used to obtain an alternative proof that $\av{q_\lambda}_S=\partial\ln
P/\partial\lambda$. Suppose we sample $N$ trajectories, of which
$N_S$ end up at state point $S$ (or a suitably defined vicinity
thereof, in a continuous state space). We have
$P(S)=\av{N_S}/N$. Then, the Malliavin property implies $\partial
P/\partial\lambda = \av{N_S\,q_\lambda}/N$, and hence, $\partial \ln
P/\partial\lambda = (\partial P/\partial\lambda)/P =
\av{N_S\,q_\lambda}/\av{N_S}=\av{q_\lambda}_S$.

\section{Multiple variables, second derivatives and the algebra of 
Malliavin weights}
Up to now, we have assumed that the quantity, $A$, does not depend
on the parameter, $\lambda$. There may be cases, however,
when $A$ does have an explicit $\lambda$-dependence. In these cases,
\Eqref{eq2} should be replaced by
\begin{equation}
\frac{\partial\av{A}}{\partial\lambda} = \mybigav{\frac{\partial
 A}{\partial\lambda}}+\av{A\,q_\lambda}\,.
\label{eq:g1}
\end{equation}
This reveals a kind of `algebra' for Malliavin weights: we see that
the operations of taking an expectation value and taking a derivative
can be commuted, provided the Malliavin weight is introduced as
the~commutator.

We can also extend our analysis further to allow us to compute higher
derivatives with respect to the parameters. These may be useful, for
example, for increasing the efficiency of gradient-based parameter
optimisation algorithms. Taking the derivative of \Eqref{eq:g1}
with respect to a second parameter,~$\mu$,~gives
\begin{align}
&\frac{\partial^2\av{A}}{\partial\lambda\partial\mu} = 
\frac{\partial}{\partial\mu}\mybigav{\frac{\partial
 A}{\partial\lambda}}+\frac{\partial\av{A\,q_\lambda}}{\partial\mu}
\nonumber\\[6pt]
&\quad=\mybigav{\frac{\partial^2\! A}{\partial\lambda\partial\mu}}
+\mybigav{\frac{\partial A}{\partial\lambda}\, q_\mu }
+\mybigav{A\,\frac{\partial q_\lambda}{\partial\mu}}\\[6pt]
&{}\hspace{9em}{}
+\mybigav{\frac{\partial A}{\partial\mu}\,q_\lambda }
+\av{A\,q_\lambda \,q_\mu}\nonumber \\[6pt]
&=\av{A\,(q_{\lambda\mu}+q_\lambda q_\mu)}
+\mybigav{\frac{\partial A}{\partial\lambda}\,q_\mu }
+\mybigav{\frac{\partial A}{\partial\mu}\,q_\lambda }
+\mybigav{\frac{\partial^2\! A}{\partial\lambda\partial\mu}}\,.\nonumber
\end{align}
In the second line, we iterate the commutation relation and, in the
third line, we collect like terms and introduce
\begin{equation}
q_{\lambda\mu}=\frac{\partial q_\lambda}{\partial\mu}\,.
\label{eq:qlmdef}
\end{equation} 
In the case where $A$ is independent of the
parameters, this result simplifies to
\begin{equation}
\frac{\partial^2\av{A}}{\partial\lambda\partial\mu} = 
\av{A\,(q_{\lambda\mu}+q_\lambda q_\mu)}\,.
\label{eq:qlm}
\end{equation}
The quantity, $q_{\lambda\mu}$, here is a new, second order Malliavin
weight, which, from \Eqsref{eq:qw} and~\eqref{eq:qlmdef}, satisfies
\begin{equation}
q_{\lambda\mu}=\frac{\partial^2\ln\curlyW}{\partial\lambda\partial\mu}\,.
\end{equation}
To compute second derivatives with respect to the parameters, we should
therefore track these second order Malliavin weights in our
simulation, updating them alongside the existing Malliavin weights by
the rule
\begin{equation}
q_{\lambda\mu}'=q_{\lambda\mu}+\frac{\partial^2\ln W(S\to S')}
{\partial\lambda\partial\mu}\,.
\end{equation}

A corollary, if we take $A$ as a constant in
\Eqsref{eq:g1} and \eqref{eq:qlm} respectively, 
is that quite generally
$\av{q_\lambda} = 0$ and
$\av{q_{\lambda\mu}}=-\av{q_\lambda q_\mu}$.

Steady state problems can be approached by extending the correlation
function method to second order weights. Define, 
\cf\ \Eqref{eq:c1},
\begin{equation}
\begin{split}
&C(t,t')=\langle A(t)\,\{[q_{\lambda\mu}(t)+q_\lambda(t) q_\mu(t)]\\[3pt]
&\hspace{9em}{}-[q_{\lambda\mu}(t')+q_\lambda(t') q_\mu(t')]\}\rangle\,.
\end{split}
\end{equation}
As in the first order case, in steady state, we expect
$C(t,t') = C(t-t')$, with the property that $C(\Delta
t)\to{\partial^2\av{A}}/{\partial\lambda\partial\mu}$ as $\Delta
t\to\infty$.

\section{One-dimensional Brownian motion in a force field}
We now demonstrate this machinery by way of a practical but very
simple example, namely one-dimensional (overdamped) Brownian motion in
a force field. In this case, the state space is specified by
the particle position, $x$, which evolves according to the Langevin
equation
\begin{equation}
\frac{dx}{dt}=f(x)+\eta\,.
\end{equation}
In this $f(x)$ is the force field and $\eta$ is Gaussian white noise of
amplitude $2T$, where $T$ is temperature. Without loss of generality
we have chosen units so that there is no prefactor multiplying the
force field. We discretise the Langevin equation to the following
updating rule
\begin{equation}
x'=x+f(x)\,\dt+\xi
\label{eq:1dbm}
\end{equation}
where $\delta t$ is the time step and $\xi$ is a Gaussian random
variate with zero mean and variance $2T\,\dt$. Corresponding to this
updating rule is an explicit expression for the propagator
\begin{equation}
W(x\to x')=\frac{1}{\sqrt{4\pi T\,\dt}}
\,\exp\Bigl(-\frac{(x'-x-f(x)\,\dt)^2}
{4T\,\dt}\Bigr)\,.
\label{eq:e}
\end{equation}
This follows from the statistical distribution of $\xi$. Let us
suppose that the parameter of interest, $\lambda$, enters into the force
field (the temperature, $T$, could also be chosen as a parameter).
Making this assumption
\begin{equation}
\frac{\partial\ln W(x\to x')}{\partial\lambda}=
\frac{(x'-x-f\,\dt)}{2T}\,\frac{\partial f}{\partial\lambda} \,.
\label{eq:e2}
\end{equation}
We can simplify this result by noting that from \Eqref{eq:1dbm},
$x'-x-f\,\dt=\xi$. Making use of this, the final updating rule for
the Malliavin weight is
\begin{equation}
q_\lambda'=q_\lambda+
\frac{\xi}{2T}\,\frac{\partial f}{\partial\lambda}
\label{eq:e3}
\end{equation}
where $\xi$ is the {\em{exact same}} value that was used for updating
the position in \Eqref{eq:1dbm}. Because the value of $\xi$ is
the same for the updates of position and of $q_\lambda$, the change in
$q_\lambda$ is completely determined by the end points, $x$ and $x'$.
The derivative, $\partial f/\partial \lambda$, should be evaluated at
$x$, since that is the position at which the force is computed in
\Eqref{eq:1dbm}. Since $\xi$ in \Eqref{eq:1dbm} is a random
variate uncorrelated with $x$, averaging \Eqref{eq:e3} shows that
$\av{q_\lambda'}=\av{q_\lambda}$. As the initial condition is
$q_\lambda=0$, this means that $\av{q_\lambda}=0$, as predicted in
the previous section. \Eqref{eq:e3} is essentially the same as
that derived in Ref.~\cite{WA12b}.

If we differentiate \Eqref{eq:e2} with respect to a second
parameter, $\mu$, we get
\begin{equation}
\begin{split}
\frac{\partial^2\ln W(x\to x')}{\partial\lambda\partial\mu}&=
\frac{(x'-x-f\,\dt)}{2T}\,\frac{\partial^2 f}{\partial\lambda\partial\mu}\\[6pt]
&\hspace{7em}{}-\frac{\dt}{2T}\,\frac{\partial f}{\partial\lambda} 
\,\frac{\partial f}{\partial\mu}\,.
\end{split}
\label{eq:e4}
\end{equation}
Hence, the updating rule for the second order Malliavin weight can be
written as
\begin{equation}
q_{\lambda\mu}'=q_{\lambda\mu}+
\frac{\xi}{2T}\,\frac{\partial^2 f}{\partial\lambda\partial\mu}
-\frac{\dt}{2T}\,\frac{\partial f}{\partial\lambda} 
\,\frac{\partial f}{\partial\mu}
\label{eq:e5}
\end{equation}
where, again, $\xi$ is the exact same value as that used for updating
the position in \Eqref{eq:1dbm}. If we average
\Eqref{eq:e5} over replicate simulation runs, we find
$\av{q_{\lambda\mu}'}=\av{q_{\lambda\mu}}-(\dt/2T)(\partial
f/\partial\lambda)(\partial f/\partial\mu)$. Hence, the mean value,
$\av{q_{\lambda\mu}}$, drifts in time, unlike $\av{q_\lambda}$ or
$\av{q_\mu}$. However, one can show that the mean value of the sum,
$\av{(q_{\lambda\mu}+q_\lambda q_\mu)}$, is constant in time and equal
to zero, as long as, initially, $q_\lambda = q_\mu = 0$.

\begin{figure}
\centering
\includegraphics[clip=true,width=0.8\columnwidth]{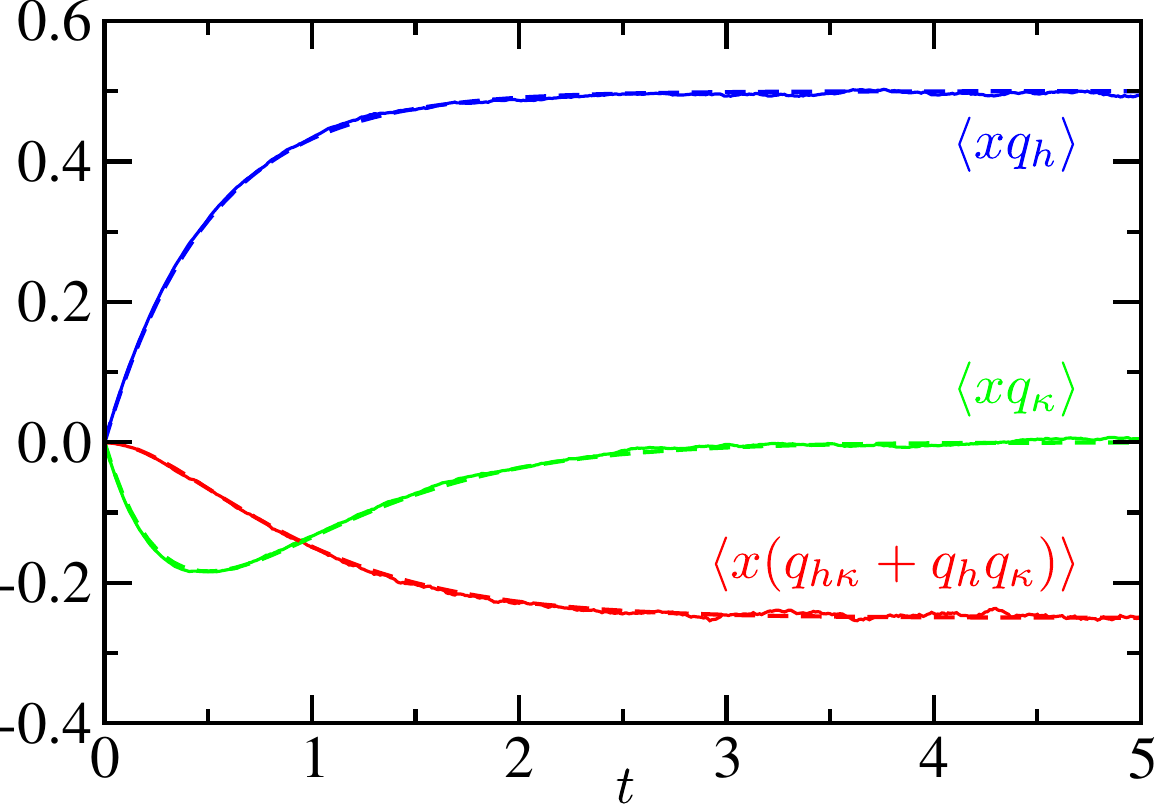}
\caption{Time-dependent derivatives, $\partial\av{x}/\partial h$ (top
  curve, blue), $\partial\av{x}/\partial\kappa$ (middle curve, green)
  and $\partial^2\av{x}/\partial h \partial\kappa$ (bottom curve,
  red). Solid lines (slightly noisy) are the Malliavin weight
  averages, generated by running the \MATLAB\ script described in
  Appendix \ref{app:script}. Dashed lines are theoretical predictions
  from \Eqsref{eq:1dtrap}.}
\label{fig1}
\end{figure}

Now, let us consider the simplest case of a particle in a linear force
field, $f=-\kappa x+h$ (also discussed in Ref.~\cite{WA12b}). This
corresponds to a harmonic trap with the potential $U=\frac{1}{2}\kappa
x^2-hx$. We let the particle start from $x_0$ at $t = 0$ and track its
time-dependent relaxation to the steady state. We shall set $T = 1$ for
simplicity. The Langevin equation can be solved exactly for this case,
and the mean position evolves according to
\begin{equation}
\av{x(t)}=x_0 e^{-\kappa t}+\frac{h}{\kappa}(1-e^{-\kappa t})\,.
\label{eq10}
\end{equation}
We suppose that we are interested in derivatives with respect to both
$h$ and $\kappa$, for a ``baseline'' parameter set in which $\kappa$ is
finite, but $h = 0$. Taking derivatives of \Eqref{eq10} and setting
$h = 0$, we find
\begin{equation}
\begin{split}
&\frac{\partial\av{x(t)}}{\partial h}=\frac{1-e^{-\kappa
 t}}{\kappa}\,,\quad
\frac{\partial\av{x}(t)}{\partial\kappa}=-x_0 t e^{-\kappa t}\,,\\[6pt]
&\frac{\partial^2\av{x(t)}}{\partial h\partial \kappa}
=\frac{t e^{-\kappa t}}{\kappa} - \frac{1-e^{-\kappa t}}{\kappa^2}\,.
\end{split}
\label{eq:1dtrap}
\end{equation}
We now show how to compute these derivatives using Malliavin weight
sampling. Applying the definitions in \Eqsref{eq:e3} and
\eqref{eq:e5}, the Malliavin weight increments are
\begin{equation}
q_h'=q_h+\frac{\xi}{2}\,,\quad
q_\kappa'=q_\kappa - \frac{x\,\xi}{2}\,,\quad
q_{h\kappa}'=q_{h\kappa}+\frac{x\,\dt}{2}\,,
\label{eq:1dq}
\end{equation}
and the position update itself is
\begin{equation}
x'=x-\kappa x\,\dt+\xi\,.
\label{eq:1dx}
\end{equation}
We track these Malliavin weights in our simulation and use them to
calculate derivatives according to
\begin{equation}
\begin{split}
&\frac{\partial\av{x(t)}}{\partial h}=\av{x(t) q_h(t)}\,,\quad
\frac{\partial\av{x(t)}}{\partial\kappa}=\av{x(t) q_\kappa(t)}\,,\\[6pt]
&\frac{\partial^2\av{x(t)}}{\partial h\partial \kappa}
=\av{x(t) (q_{h\kappa}(t)+q_h(t) q_\kappa(t))}\,.
\end{split}
\label{eq:mws2}
\end{equation}
\Eqsref{eq:1dq}--\eqref{eq:mws2} have been coded up as a
\MATLAB\ script, described in Appendix \ref{app:script}. A typical result
generated by running this script is shown in
Fig.~\ref{fig1}.  \Eqsref{eq:1dq} and \eqref{eq:1dx} are
iterated with $\dt = 0.01$ up to $t = 5$, for a trap strength $\kappa
= 2$ and initial position $x_0 = 1$. The weighted averages in
\Eqref{eq:mws2} are evaluated as a function of time, for $N =
10^5$ samples, as in \Eqref{eq:samp}. These results are shown
as the solid lines in Fig.~\ref{fig1}. The dashed lines are
theoretical predictions for the time dependent derivatives from
\Eqsref{eq:1dtrap}. As can be seen, the agreement between the
time-dependent derivatives and the Malliavin weight averages is very
good.

\begin{figure}
\centering
\includegraphics[clip=true,width=0.8\columnwidth]{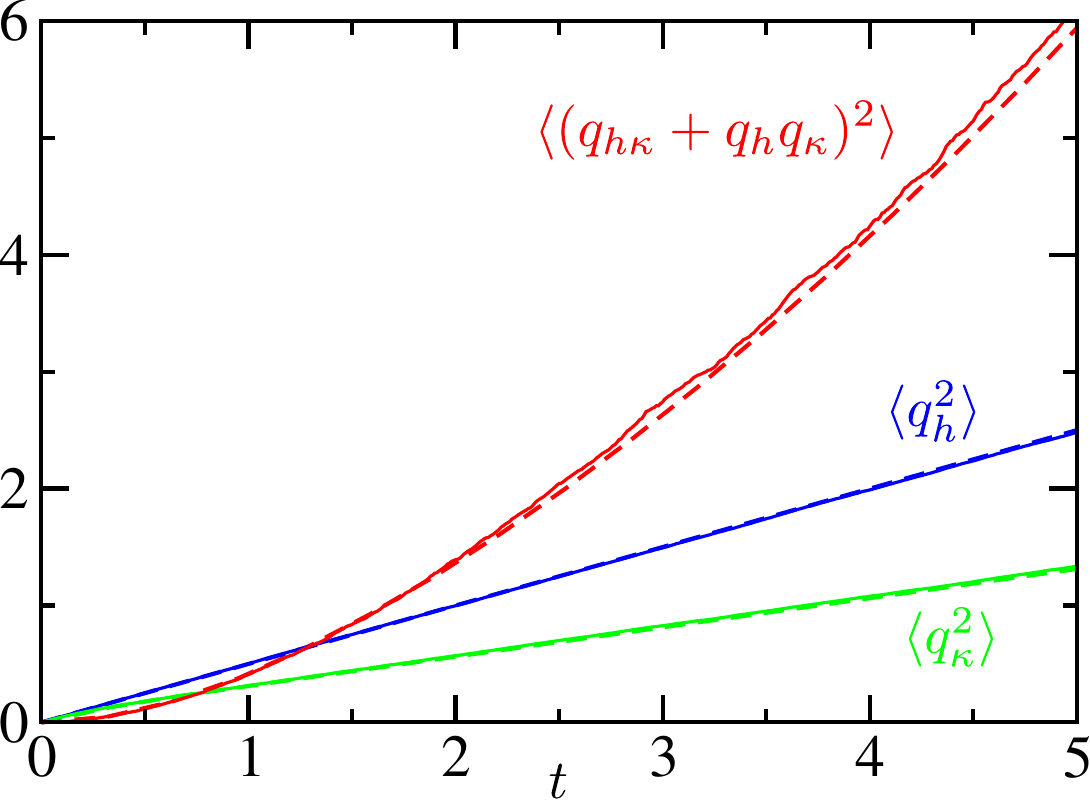}
\caption{Growth of mean square Malliavin weights with time. The solid
  lines are from simulations and the dashed lines are from
  \Eqsref{eq:app1} in the Appendix. Parameters are as for
  Fig.~\ref{fig1}.}
\label{fig2}
\end{figure}

As discussed briefly above, in this procedure, the sampling error in
the computation of $\partial \av{A(t)}/\partial \lambda$ is expected
to grow with time. Fig.~\ref{fig2} shows the mean square Malliavin
weight as a function of time for the same problem. For the first
order weights, $q_h$ and $q_\kappa$, the growth rate is typically linear
in time. Indeed, from \Eqsref{eq:1dq}, one can prove that in the
limit $\dt\to0$ (see Appendix \ref{app:anal})
\begin{equation}
\frac{d\av{q_h^2}}{dt}=\frac{1}{2}\,,\quad
\frac{d\av{q_\kappa^2}}{dt}=\frac{\av{x^2}}{2}\,.
\end{equation}
Thus $q_h$ behaves exactly as a random walk, as should be obvious from
the updating rule. The other weight, $q_\kappa$, also ultimately
behaves as a random walk, since $\av{x^2}=1/\kappa$ in steady state
(from equipartition). Fig.~\ref{fig2} also shows that the second
order weight, $q_{h\kappa}$, grows superdiffusively; one can show that,
eventually, $\av{(q_{h\kappa}+q_h q_\kappa)^2}\sim t^2$, although the
transient behaviour is complicated. Full expressions are given in
Appendix \ref{app:anal}. This suggests that computation of second order
derivatives is likely to suffer more severely from statistical
sampling problems than the computation of first order derivatives.

\section{Conclusions}
In this paper, we have provided an outline of the generic use of
Malliavin weights for sampling derivatives in stochastic simulations,
with an emphasis on practical aspects. The usefulness of MWS for a
particular simulation scheme hinges on the simplicity, or otherwise,
of constructing the propagator, $W(S\to S')$, which fixes the updating
rule for the Malliavin weights according to \Eqref{eq:prop4}. The
propagator is determined by the algorithm used to implement the
stochastic equations of motion; MWS may be easier to implement for
some algorithms than for others. We note, however, that there is
often some freedom of choice about the algorithm, such as the choice
of a stochastic thermostat in molecular dynamics, or the order in
which update steps are implemented. In these cases, a suitable choice
may simplify the construction of the propagator and facilitate the use
of Malliavin weights.
\begin{center}
---oOo---
\end{center}
Rosalind J. Allen is supported by a Royal Society University Research
Fellowship.

\appendix

\section{Selected analytic results}\label{app:anal}
Here, we present analytic results for the growth in time of the mean
square Malliavin weights. We can express the rate of growth of the
mean of a generic function, $f(x,q_h,q_\kappa,q_{h\kappa})$, as
\begin{equation}
\frac{d\av{f}}{dt}=\lim_{\dt\to0}
\frac{\av{f(x',q_h',q_\kappa',q_{h\kappa}')
-f(x,q_h,q_\kappa,q_{h\kappa})}}{\delta t}\,.
\end{equation}
On the right-hand side (RHS), the values of $x'$, $q_h'$,
$q_\kappa'$ and $q_{h\kappa}$ are substituted from the updating rules
in \Eqsref{eq:1dq} and \eqref{eq:1dx}. In calculating the RHS
average, we note that the distribution of $\xi$ is a Gaussian
independent of the position and Malliavin weights, and thus, one can
substitute $\av{\xi} = 0$, $\av{\xi^2} = 2\,\dt$, $\av{\xi^3} = 0$,
$\av{\xi^4} = 12\,\dt^2$, \etc. Proceeding in this way, with
judicious choices for $f$, one can obtain the following set of coupled
ordinary differential equations (ODEs)
\begin{align}
&\frac{d\av{q_h^2}}{dt}=\frac{1}{2}\,,\quad
\frac{d\av{q_\kappa^2}}{dt}=\frac{\av{x^2}}{2}\,,\nonumber\\[9pt]
&\frac{d\av{x^2}}{dt}+2\kappa\av{x^2}=2\,,\quad
\frac{d\av{x q_h}}{dt}+\kappa\av{x q_h}=1\,,\nonumber\\[9pt]
&\frac{d\av{x^2q_h^2}}{dt}+2\kappa\av{x^2q_h^2}
=2\av{q_h^2}+4\av{x q_h}+\frac{\av{x^2}}{2}\,,\nonumber\\[9pt]
&\frac{d\av{xq_hq_\kappa}}{dt}+\kappa\av{xq_hq_\kappa}
=-\av{x q_h}-\frac{\av{x^2}}{2}\,,\\[9pt]
&\frac{d\av{(q_{h\kappa}+q_hq_\kappa)^2}}{dt}
=\frac{\av{q_\kappa^2}}{2}-\av{x q_hq_\kappa}
+\frac{\av{x^2q_h^2}}{2}\nonumber\\[6pt]
&\hspace{9em}\Bigl({}=\frac{\av{(q_\kappa-x q_h)^2}}{2}\Bigr)\,.\nonumber
\end{align}
Some of these have already been encountered in the main text. The
last one is for the desired mean square second order weight. The ODEs
can be solved with the initial conditions that at $t = 0$, all averages
involving Malliavin weights vanish, but $\av{x^2} = x_0^2$. The
results include \interalia\
\begin{align}
&\av{q_h^2}=\frac{t}{2}\,,\quad
\av{q_\kappa^2}=\frac{t}{2\kappa}
+\frac{(\kappa x_0^2-1)(1-e^{-2\kappa t})}{4\kappa^2}\,,\nonumber\\[9pt]
&\av{(q_{h\kappa}+q_hq_\kappa)^2}=
\frac{2\kappa^2t^2+(19+\kappa x_0^2)\kappa t+2\kappa x_0^2-34}
{8\kappa^3}\nonumber\\[9pt]
&\hspace{6em}{}+\frac{2\kappa t+10-\kappa x_0^2}{2\kappa^3}
\,e^{-\kappa t}\nonumber\\[6pt]
&\hspace{9em}{}+
\frac{(1-\kappa x_0^2)\kappa t+2\kappa x_0^2-6}{8\kappa^3}\,e^{-2\kappa t}\,.
\label{eq:app1}
\end{align}
These are shown as the dashed lines in Fig.~\ref{fig2}. The leading
behaviour of the last as $t\to\infty$ is
\begin{equation}
\av{(q_{h\kappa}+q_hq_\kappa)^2}=\frac{t^2}{4\kappa}
+\text{subdominant terms}\,.
\end{equation}
However, the approach to this limit is slow.

\section{MATLAB script}\label{app:script}
The \MATLAB\ script in Listing \ref{list1} was used to generate the
results shown in Fig.~\ref{fig1}. It implements
\Eqsref{eq:1dq}--\eqref{eq:mws2} above, making extensive use of
the compact \MATLAB\ syntax for array operations, for instance, invoking
`{\ttfamily .*}' for element-by-element multiplication of arrays.

Here is a brief explanation of the script. \LineRange{1}{3}
initialise the problem and the parameter values. \Lines{4}{5}
calculate the number of points in a trajectory and initialise a vector
containing the time coordinate of each point.
\LineRange{6}{9} set aside storage for the actual trajectory,
Malliavin weights and cumulative statistics. \LineRange{10}{23}
implement a pair of nested loops, which are the kernel of the
simulation. Within the outer (trajectory sampling) loop, \Line{11}
initialises the particle position and Malliavin weights, \Line{12}
precomputes a vector of random displacements (Gaussian random
variates) and \LineRange{13}{18} generate the actual trajectory.
Within the inner (trajectory generating loop), \LineRange{14}{17} are a
direct implementation of \Eqsref{eq:1dq} and \eqref{eq:1dx}.
After each individual trajectory has been generated, the cumulative
sampling step implied by \Eqref{eq:mws2} is done in
\LineRange{19}{22}; after all the trajectories have been generated,
these quantities are normalised in \Lines{24}{25}. Finally,
\LineRange{26}{32} generate a plot similar to Fig.~\ref{fig1} (albeit with
the addition of $\av{x}$), and \Lines{33}{34} show how the data can
be exported in tabular format for replotting using an external
package.

Listing \ref{list1} is complete and self-contained. It will run in
either \MATLAB\ or \Octave. One minor comment is perhaps in order. The
choice was made to precompute a vector of Gaussian random variates,
which are used as random displacements to generate the trajectory and
update the Malliavin weights. One could equally well generate random
displacements on-the-fly, in the inner loop. For this one-dimensional
problem, storage is not an issue, and it seems more elegant and
efficient to exploit the vectorisation capabilities of \MATLAB. For a
more realistic three-dimensional problem, with many particles (and a
different programming language), it is obviously preferable to use an
on-the-fly approach.\\[0.2in]

%\bibliography{malliavin}

\begin{thebibliography}{9}
\expandafter\ifx\csname natexlab\endcsname\relax\def\natexlab#1{#1}\fi
\expandafter\ifx\csname bibnamefont\endcsname\relax
  \def\bibnamefont#1{#1}\fi
\expandafter\ifx\csname bibfnamefont\endcsname\relax
  \def\bibfnamefont#1{#1}\fi
\expandafter\ifx\csname citenamefont\endcsname\relax
  \def\citenamefont#1{#1}\fi
\expandafter\ifx\csname url\endcsname\relax
  \def\url#1{\texttt{#1}}\fi
\expandafter\ifx\csname urlprefix\endcsname\relax\def\urlprefix{URL }\fi
\providecommand{\bibinfo}[2]{#2}
\providecommand{\eprint}[2][]{\url{#2}}

\bibitem[{\citenamefont{Bell}(2006)}]{Bel07}
\bibinfo{author}{\bibfnamefont{D.~R.} \bibnamefont{Bell}},
  \emph{\bibinfo{title}{The {M}alliavin calculus}} (\bibinfo{publisher}{Dover},
  \bibinfo{address}{Mineola, New York}, \bibinfo{year}{2006}).

\bibitem[{\citenamefont{Nualart}(2006)}]{Nua06}
\bibinfo{author}{\bibfnamefont{D.}~\bibnamefont{Nualart}},
  \emph{\bibinfo{title}{The {M}alliavin calculus and related topics}}
  (\bibinfo{publisher}{Springer-Verlag}, \bibinfo{address}{New York},
  \bibinfo{year}{2006}).

\bibitem[{\citenamefont{Fourni{\'e} et~al.}(1999)\citenamefont{Fourni{\'e},
  Lasry, Lebuchoux, Lions, and Touzi}}]{FLL+99}
\bibinfo{author}{\bibfnamefont{E.}~\bibnamefont{Fourni{\'e}}},
  \bibinfo{author}{\bibfnamefont{J.-M.} \bibnamefont{Lasry}},
  \bibinfo{author}{\bibfnamefont{J.}~\bibnamefont{Lebuchoux}},
  \bibinfo{author}{\bibfnamefont{P.-L.} \bibnamefont{Lions}}, \bibnamefont{and}
  \bibinfo{author}{\bibfnamefont{N.}~\bibnamefont{Touzi}},
  \bibinfo{journal}{Finance Stochast.} \textbf{\bibinfo{volume}{3}},
  \bibinfo{pages}{391} (\bibinfo{year}{1999}).

\bibitem[{\citenamefont{Plyasunov and Arkin}(2007)}]{PA07}
\bibinfo{author}{\bibfnamefont{A.}~\bibnamefont{Plyasunov}} \bibnamefont{and}
  \bibinfo{author}{\bibfnamefont{A.~P.} \bibnamefont{Arkin}},
  \bibinfo{journal}{J. Comp. Phys.} \textbf{\bibinfo{volume}{221}},
  \bibinfo{pages}{724} (\bibinfo{year}{2007}).

\bibitem[{\citenamefont{Berthier}(2007)}]{Ber07}
\bibinfo{author}{\bibfnamefont{L.}~\bibnamefont{Berthier}},
  \bibinfo{journal}{Phys. Rev. Lett.} \textbf{\bibinfo{volume}{98}},
  \bibinfo{pages}{220601} (\bibinfo{year}{2007}).

\bibitem[{\citenamefont{Chen and Glasserman}(2007)}]{CG07}
\bibinfo{author}{\bibfnamefont{N.}~\bibnamefont{Chen}} \bibnamefont{and}
  \bibinfo{author}{\bibfnamefont{P.}~\bibnamefont{Glasserman}},
  \bibinfo{journal}{Stoch. Proc. Appl.} \textbf{\bibinfo{volume}{117}},
  \bibinfo{pages}{1689} (\bibinfo{year}{2007}).

\bibitem[{\citenamefont{Warren and Allen}(2012{\natexlab{a}})}]{WA12}
\bibinfo{author}{\bibfnamefont{P.~B.} \bibnamefont{Warren}} \bibnamefont{and}
  \bibinfo{author}{\bibfnamefont{R.~J.} \bibnamefont{Allen}},
  \bibinfo{journal}{J. Chem. Phys.} \textbf{\bibinfo{volume}{136}},
  \bibinfo{pages}{104106} (\bibinfo{year}{2012}{\natexlab{a}}).

\bibitem[{\citenamefont{Warren and Allen}(2012{\natexlab{b}})}]{WA12b}
\bibinfo{author}{\bibfnamefont{P.~B.} \bibnamefont{Warren}} \bibnamefont{and}
  \bibinfo{author}{\bibfnamefont{R.~J.} \bibnamefont{Allen}},
  \bibinfo{journal}{Phys. Rev. Lett.} \textbf{\bibinfo{volume}{109}},
  \bibinfo{pages}{250601} (\bibinfo{year}{2012}{\natexlab{b}}).

\bibitem[{\citenamefont{Warren and {ten Wolde}}(2005)}]{WtW05}
\bibinfo{author}{\bibfnamefont{P.~B.} \bibnamefont{Warren}} \bibnamefont{and}
  \bibinfo{author}{\bibfnamefont{P.~R.} \bibnamefont{{ten Wolde}}},
  \bibinfo{journal}{J. Phys. Chem. B} \textbf{\bibinfo{volume}{109}},
  \bibinfo{pages}{6812} (\bibinfo{year}{2005}).

\end{thebibliography}

\onecolumngrid

\lstset{language=Matlab}
\lstset{numbers=left, numberstyle=\tiny, stepnumber=1, numbersep=5pt, %
 keywordstyle=\bfseries, basicstyle=\ttfamily, frame=lines}

\begin{lstlisting}[float, upquote=true,
 caption={\MATLAB\ script used to generate Figure 1.}, label=list1]
clear all
randn('seed', 12345);
kappa = 2; x0 = 1; tend = 5; dt = 0.01; nsamp = 10^5;
npt = round(tend/dt) + 1;
t = (0:npt-1)' * dt;
x = zeros(npt, 1); xi = zeros(npt, 1);
qh = zeros(npt, 1); qk = zeros(npt, 1); qhk = zeros(npt, 1); 
x_av = zeros(npt, 1); xqh_av = zeros(npt, 1); 
xqk_av = zeros(npt, 1); xqhk_av = zeros(npt, 1);
for samp = 1:nsamp
  x(1) = x0; qh(1) = 0; qk(1) = 0; qhk(1) = 0;
  xi = randn(npt, 1) * sqrt(2*dt);
  for i = 1:npt-1
    x(i+1) = x(i) - kappa*x(i)*dt + xi(i);
    qh(i+1) = qh(i) + 0.5*xi(i);
    qk(i+1) = qk(i) - 0.5*x(i)*xi(i);
    qhk(i+1) = qhk(i) + 0.5*x(i)*dt;
  end
  x_av = x_av + x;
  xqh_av = xqh_av + x.*qh;
  xqk_av = xqk_av + x.*qk;
  xqhk_av = xqhk_av + x.*(qhk + qh.*qk);
end
x_av = x_av / nsamp; xqh_av = xqh_av / nsamp;
xqk_av = xqk_av / nsamp; xqhk_av = xqhk_av / nsamp;
hold on
plot(t, x_av, 'k'); plot(t, xqh_av, 'b')
plot(t, xqk_av, 'g'); plot(t, xqhk_av, 'r')
plot(t, x0*exp(-kappa*t), 'k--')
plot(t, (1-exp(-kappa*t))/kappa, 'b--')
plot(t, -x0*t.*exp(-kappa*t), 'g--')
plot(t, t.*exp(-kappa*t)/kappa-(1-exp(-kappa*t))/(kappa^2), 'r--')
result = [t x_av xqh_av xqk_av xqhk_av];
save('result.dat', '-ascii', 'result')
\end{lstlisting}

\twocolumngrid

\end{document}